# A Flexible and Secure Remote Systems Authentication Scheme Using Smart Cards

Manik Lal Das

*Dhirubhai Ambani Institute of Information and Communication Technology*
*Gandhinagar – 382007, India.*
*Email: maniklal_das@daiict.ac.in*

**Abstract** — The paper presents an authentication scheme for remote systems using smart card. The scheme prevents the scenario of *many* logged in users with the *same* login identity, and does not require password/verifier table to validate the users' login request. The scheme provides a user-friendly password change option, and withstands the replay, impersonation, stolen-verifier, guessing, and denial-of-service attacks[1].

## I. INTRODUCTION

Remote system authentication is a process by which a remote system gains confidence about the identity of the communicating partner. In 1981, Lamport [1] first introduced the concept of password-based authentication scheme. Afterwards, numerous studies of remote systems authentication with emphases on various types of mathematical problems (e.g. one-way hash function, public-key setting) have been investigated to improve the security and efficiency of the scheme. Taking computational cost into consideration, remote systems authentication can be classified into two broad categories: password-based using one-way hash function and public key [2] based techniques. One-way hash function based authentication technique is simple and viable for implementation in a small handheld device like smart card. In contrast, the public-key based technique provides higher security with added computational cost. It is observed that Lamport's scheme suffers high hash overhead and the necessity for password resetting problems decreases its suitability for practical use. Haller [3] proposed a modified version of Lamport's scheme, but the modified version is vulnerable to the replay attack. Shimizu [4] proposed a one-time password authentication scheme on eliminating the weaknesses of [1] and [3]. The one-time characteristic is gained by using two variable random numbers that are changed in every authentication. The user has to memorize two variable random numbers or carry with some sort of portable storage token, e.g. smart card. In 2000, Sandirigama et al. [5] proposed another hash-based strong-password authentication scheme, but the scheme is vulnerable [6] to some security threats. Later Peyravian and Zunic [7] proposed a hash-based password authentication scheme that requires low computational efforts. However, Hwang and Yeh [8] showed that Peyravian-Zunic's scheme is vulnerable to password guessing, server spoofing, and stolen-verifier attacks. In 2004, Das et al. proposed a scheme [9] that provides computational efficiency and a user-friendly of password change option. Subsequently, a few public-key based remote systems authentication schemes and improvements [10], [11], [12], [13] have also been proposed. We observe that all these schemes provide only one-way authentication, that is, only remote server can check the authenticity of a user. The user cannot check whether he is communicating with the correct server or not. It is a vital gap where a potential adversary can spoof the server and get valuable user information. This motivates us to construct an authentication scheme for remote systems that provides user and server authentication, and the user gets access to the remote system's resource only if <user, server>'s authenticity passed correctly.

To construct the scheme, we use a smart card as the user registration token and the merits of "password" and "smart card" factors for the remote systems authentication. The use of smart card not only make the scheme secure but also prevents users from distribution of their login identities, which effectively prohibits the scenario of *many* logged in users with the *same* login-identity (login-ID). This generally happens in digital libraries and similar systems, where a subscriber can share his login-ID and password with others. The scheme is computationally efficient and the remote system does not need to maintain any password/verifier table for users login request validation. Furthermore, the users are allowed to choose and change their preferred passwords without the remote systems assistance. The rest of the paper is organized as follows: In Section 2, we discuss a widely accepted list of security properties for remote systems authentication. We present the scheme in Section 3. The security and efficiency of the scheme are analyzed in Section 4. Finally, we conclude the paper in Section 5.

---

## II. SECURITY PROPERTIES

Desirable security properties of various remote systems authentication have been evolved over this period. A widely accepted list of required properties is given below:

- *Replay Attack*: Replay attack is an offensive action whereby an adversary can intercept a valid login message and then gains access to the remote system by replaying the intercepted information.
- *Impersonation Attack*: An adversary may impersonate user login by forging a login request and acts as a legitimate user by logging in to the server.
- *Stolen-verifier Attack*: In practice, it is likely that a user uses the same password to access several servers for his convenience. If an insider of a remote server obtains user password, he could impersonate user's login to access other servers. Generally, the system, which maintains verifier/password table for user's login request verification, may suffer from this attack.
- *Guessing Attack*: Memorized passwords are subject to guessing attacks. The guessing attacks can be classified into two types: on-line password guessing attack and off-line password guessing attack.
  a. On-line password guessing attack: The adversary can try to use a guessed password to pass the verification of remote server in on-line manner. The server could detect the attacks by noticing continuous authentication failures.
  b. Off-line password guessing attack: The adversary can intercept a valid login request and store them locally. Then he would try to find out the password in trial and error method, and verify the login request in off-line manners. Therefore, the remote server could not detect the attacks.
- *Denial-of-Service Attack*: A denial-of-service attack is an offensive action whereby the adversary could use some method to work upon the remote server so that the server will deny the access requests issued by the legitimate user.

## III. PROPOSED SCHEME

The scheme consists of three phases: registration phase, authentication phase and password change phase. The registration phase is performed only once, and the authentication phase is executed every time when a user wants to login to the system.

### A. Registration Phase

This phase is invoked whenever a user $U_i$ wants to register to the remote system (RS). The user chooses a password $PW_i$ and submits it to the RS. Upon receiving the registration request, the RS performs the following steps:

R1. Compute a nonce $N_i = h(PW_i, ID_i) \oplus h(x)$, where $x$ is a primary secret key of RS, h(.) is a one-way hash function and $\oplus$ is a bitwise concatenation operator.

R2. Personalize a smart card with the parameters h(.), $ID_i$, $N_i$, $h(PW_i)$ and $y$. The parameter $y$ is RS's secondary secret number stored in each registered user's smart card. Then, the RS sends the personalized smart card to $U_i$ in a secure manner.

### B. Authentication Phase

This phase is invoked whenever $U_i$ wants to login to the RS. The phase is further divided into two parts, namely the *User authentication* and *RS authentication*. The phases work as follows:

*User Authentication*

$U_i$ inserts his smart card to a terminal, and keys his identity $ID_i$ and password $PW_i$. The smart card validates the entered $ID_i$ and $PW_i$ with the stored ones in smart card. If the entered $ID_i$ and $PW_i$ are correct, the smart card performs the following operations:

U1. Compute $DID_i = h(PW_i, ID_i) \oplus h(y \oplus T_u)$, where $T_u$ is timestamp of $U_i$'s system.

U2. Compute $C_i = h(N_i \oplus T_u \oplus y)$.

U3. Send $(DID_i, C_i, T_u)$ as a login request to the RS over a public channel.

Upon receiving the login request $(DID_i, C_i, T_u)$ at time $T_s$, the RS authenticates the $U_i$ by the following steps:

V1. Verify the validity of the time interval between $T_u$ and $T_s$. If $(T_s - T_u) \leq \Delta T$ then the remote system proceeds to the next step, otherwise terminates the request. The $\Delta T$ denotes the expected valid time interval for the transmission delay.

V2. Compute $h(PW_i, ID_i) = DID_i \oplus h(y \oplus T)$.

V3. Compute $C_i^* = h(h(PW_i, ID_i) \oplus h(x) \oplus T_u \oplus y)$.

V4. If $C_i^* = C_i$, the RS accepts the login request, rejects otherwise.

*RS Authentication*

If the user authenticity is passed correctly, the RS's genuineness is ascertained by the following steps:

S1. Compute $X_i = h(h(PW_i, ID_i) \oplus h(x) \oplus T_u \oplus T_s^*)$; $T_s^*$ is timestamp of RS's system.

S2. Send $(X_i, T_s^*)$ to the user over a public channel.

Let the user receives the response at time $T_u^*$. Then the smart card validates the time interval between $T_u^*$ and $T_s^*$. If $(T_u^* - T_s^*) \leq \Delta T$, it computes $X_i^* = h(N_i \oplus T_u \oplus T_s^*)$. If $X_i^* = X_i$, then the RS is authentic and $U_i$ starts accessing the RS's resource.

## C. Password Change Phase

This phase is invoked whenever $U_i$ wants to change his password. He can easily change his password without taking any assistance from the RS. The phase works as follows:

- P1. $U_i$ inserts the smart card into a terminal and submits $ID_i$ and $PW_i$ and requests to change the password. The smart card validates the entered $ID_i$ and $PW_i$ with the one that is stored in the smart card. If entered parameters are correct, the smart card proceeds to the next step, otherwise terminates the operation.

- P2. $U_i$ is prompted to submit a new password, and he submits $PW_i^*$.

- P3. The smart card computes

$$N_i^* = N_i \oplus h(PW_i, ID_i) \oplus h(PW_i^*, ID_i).$$

- P4. The nonce $N_i$ and $PW_i$ will be replaced by $N_i^*$ and $PW_i^*$ respectively, and it completes the password change phase.

## IV. ANALYSIS OF THE SCHEME

### A. Security Analysis

The replication or extraction of parameter from the private space of smart card is quite difficult as per the present literature. Though it happens by the side channel attacks [14]; however, the experiment cost is much higher than the cost of the intended parameter. Further, some of the smart card manufacturers consider the risk of the side channel attack, and provide counter measure to deter the reverse engineering attempt. We consider a smart as a secure device. With this fact, the proposed scheme withstands the following possible threats:

*Replay Attack:* A replay attack (replaying an intercepted login message) cannot work in the scheme. Suppose an attacker intercepts a valid login request ($DID$, $C$, $T_u$) and tries to login to the RS by replaying the same. The verification of this login request will fail because the interval ($T_s - T_u$) $\geq \Delta T$, a mutually agreed transmission delay. Of course, the clock synchronization at client and server needs an attention of this scheme. We believe that this is an efficient way to resist the replay attack, and assume that clock synchronization is being taking into account for the proposed scheme.

*Impersonation Attack:* An attacker cannot impersonate a legal login by intercepting ($DID$, $C$, $T_u$). The attacker will have $DID$, but this $DID$ needs to be recomputed with a new timestamp $T_{new}$, which is not possible, as it requires $PW$ and $y$. To get $PW$ and/or $y$ from $DID$ one has to break the one-wayness property of $h(.)$, which is a hard problem. Now let us see, the capability of a valid user to forge a login request. In this case, the user knows $PW$, but obtaining $y$ is again a hard problem. Thus, no one (including valid users) can forge a login request.

*Stolen-verifier Attack:* In practice, it is likely that $U$ uses the same password $PW$ to access several servers for his convenience. If an insider of the RS obtains $PW$, he could frame $U$'s login to access other servers. Our scheme resists this attack because the insider should need $U$'s smart card to frame $U$. In our scheme, $U$ initially submits $PW$ to the RS during the registration process; however, after registration $U$ is free to change his password any time. Further, the RS does not maintain any verifier/password table to validate $U$'s login request. Thus, there is no question to steal password and thereby execution of the stolen-verifier attack is not possible.

*Guessing Attack:* The guessing attack is a crucial concern in any authentication scheme. We note that our scheme is free from password/verifier table, and user password is not traveled as a hash of password. Instead, we let password to travel as a digest of some other secret components. Therefore, our scheme is not suffered by the guessing attacks.

*Denial-of-Service Attack*: In the password change phase, the denial-of-service attack may occur when the RS updates the new verifier or password for the next login without checking the validation of the entered input. This allows the RS to reject all subsequent login requests of a legal user. Therefore, it is intuitive to check the any request/input before updating the verifier or password. In our scheme, when a user logs into RS or wants to change his password, the smart card checks the validity of the card owner password before processing any instruction. Even, if a smart card is stolen or lost, the party who has the smart card cannot login to RS or change the password without knowing the card owner's password. Therefore, denial-of-service attack is not possible in our scheme.

### B. Efficiency

The smart card personalization cost for the registration process of our scheme is at per the schemes in [10], [11], [12], [13]. The login and verification phases of these schemes require multiple modular arithmetic and hash computation; whereas the login and verification phases require only hash computation. Therefore, our scheme is computationally efficient in comparisons to the schemes [10], [11], [12], [13]. As most of the hash-based schemes [1]-[5], [7], [9] suffer from the security weakness and our scheme attains its security strength against the possible threats, the proposed scheme is practical and viable in smart card based authentication schemes.

## V. CONCLUSION

We have proposed an efficient remote systems authentication scheme that provides the following characteristics:
- the scheme provides mutual authentication between user and remote server.
- the scheme prevents the scenario of many logged in users with the same login identity.

- the scheme provides a flexible password change option, where users can change their passwords any time without any assistance of remote server.
- the remote server does not require to maintain any verifier/password table to validate the login request.
- the scheme withstands the replay, impersonation, stolen-verifier, guessing, and denial-of-service attacks.

In future, the authors will try to avoid the secure channel in the registration phase so that the proposed construction would be purely public channel based remote systems authentication, and this is an interesting and challenging extension of the proposed work.